\documentclass[sigconf]{acmart}
\AtBeginDocument{%
  }

\copyrightyear{2026}
\acmYear{2026}
\setcopyright{cc}
\setcctype{by}
\acmConference[SIGCSE TS 2026]{Proceedings of the 57th ACM Technical Symposium on Computer Science Education V.2}{February 18--21, 2026}{St. Louis, MO, USA}
\acmBooktitle{Proceedings of the 57th ACM Technical Symposium on Computer Science Education V.2 (SIGCSE TS 2026), February 18--21, 2026, St. Louis, MO, USA}
\acmPrice{}
\acmDOI{10.1145/3770761.3777340}
\acmISBN{979-8-4007-2255-4/2026/02}

\acmSubmissionID{V2pp1290}

\usepackage{enumitem}
\usepackage{cuted}

\usepackage{caption} 
\usepackage{graphicx}

\begin{document}

\title{DiverseClaire: 
Simulating
Students to Improve Introductory
Programming 
Course Materials for All CS1 Learners}


 \author{Wendy Wong}
 \email{wendy.wong@student.unsw.edu.au}
 \orcid{0009-0008-2438-0075}
 \affiliation{%
 \institution{The University of New South Wales}
 \city{Sydney}
  \state{NSW}
  \country{Australia}
 }
 \author{Yuchao Jiang}
 \email{yuchao.jiang@unsw.edu.au}
 \orcid{0000-0003-2031-4712}
 \affiliation{%
 \institution{The University of New South Wales}
   \city{Sydney}
   \state{NSW}
   \country{Australia}}

 \author{Yuekang Li}
 \email{yuekang.li@unsw.edu.au}
 \orcid{0000-0003-4382-0757}
 \affiliation{%
   \institution{The University of New South Wales}
   \city{Sydney}
   \state{NSW}
   \country{Australia}
}



\renewcommand{\shortauthors}{Wendy Wong, Yuchao Jiang, \& Yuekang Li}


\begin{abstract}


Although CS programs are booming, 
introductory courses like CS1 still adopt 
a one-size-fits-all formats that can 
exacerbate cognitive load and discourage 
learners with  autism, ADHD, dyslexia and 
other neurological conditions. These call 
for compassionate pedagogies and 
Universal Design For Learning (UDL) to 
create learning environments and 
materials where cognitive diversity is 
welcomed. To address this, we introduce 
DiverseClaire a pilot study, which 
simulates students including neurodiverse 
profiles using LLMs and diverse personas. 
By leveraging Bloom’s Taxonomy and UDL, 
DiverseClaire compared UDL-transformed 
lecture slides with traditional formats. 
To evaluate  DiverseClaire controlled 
experiments, we used the evaluation 
metric the average score. The findings 
revealed that the simulated neurodiverse 
students struggled with learning due to 
lecture slides that were in inaccessible 
formats. These results highlight the need 
to provide course materials in multiple 
formats for diverse learner preferences. 
Data from our pilot study will be made 
available to assist future CS1 
instructors.


\end{abstract}

\ccsdesc[500]{Social and professional topics~Computing education programs}

\keywords{CS1, Inclusive Education, Universal Design for Learning, Large Language Models}



\maketitle

\section{Introduction}

Neurodivergent learners,
such as students with attention-deficit hyperactivity disorder 
(ADHD) or autism spectrum disorder (ASD),
enrol in undergraduate 
computer science programs 
in substantial numbers \cite{Bonnette2022BuildingNeeds}
Self-reported disabilities, 
including ASD and ADHD, 
in Australian higher education
increased by 163\% 
from 4,054 students in 2021
to 10,665 students in 2024 \footnote{https://shorturl.at/zeiYp}.
The course format and content
impact the learning experience of neurodiverse
university students, contributing to their lower completion
rates\cite{Wang2024ExploringCourses}.

%
 %
%
Additionally, course materials often
lack digital accessibility features
needed by neurodiverse students.

Universal Design For Learning (UDL) offers a framework to make teaching more inclusive. Its principle of Multiple Means
of Representation~\cite{Garrad2022RethinkingStudies} encourages educators
to present ideas and information through varied and flexible formats so that learners can access, engage with and comprehend content regardless of their sensory, linguistic or cultural needs
\cite{BeckWells2022SupportingLearning}.

Simulating students via large language models (LLMs) offers an inexpensive way to pilot pedagogical interventions \cite{Broska2025TheObservations}. By creating synthetic personas with demographic and behavioural attributes, researchers can explore how learners with diverse prior knowledge, engagement levels and learning attitudes might respond to new materials before conducting resource‑intensive human‑subject studies. 
To our knowledge, this approach has not yet been used to evaluate whether UDL‑inspired improvements to lecture slides make introductory computer science courses more accessible.


The main contributions of this work are:
\begin{itemize}[leftmargin=*]
\item LLM‑based simulations: We use RCTs with large‑language models (via few‑shot prompting and retrieval‑augmented generation) to predict student performance on UDL‑enhanced and original lecture slides.

\item Diverse personas: Our synthetic learners integrate demographic, behavioural and learning traits, enabling evaluation across a broad spectrum of neurodiverse experiences.

\item ASD insights: The simulations reveal that UDL improvements on lecture slides provided no benefit to ASD learners, indicating a need for more tailored accessibility strategies.

\end{itemize}

\vspace{-3mm}

\section{DiverseClaire}

\begin{figure*}[!t]
\vspace{-5mm}
    \centering
    \includegraphics[width=0.85\textwidth]{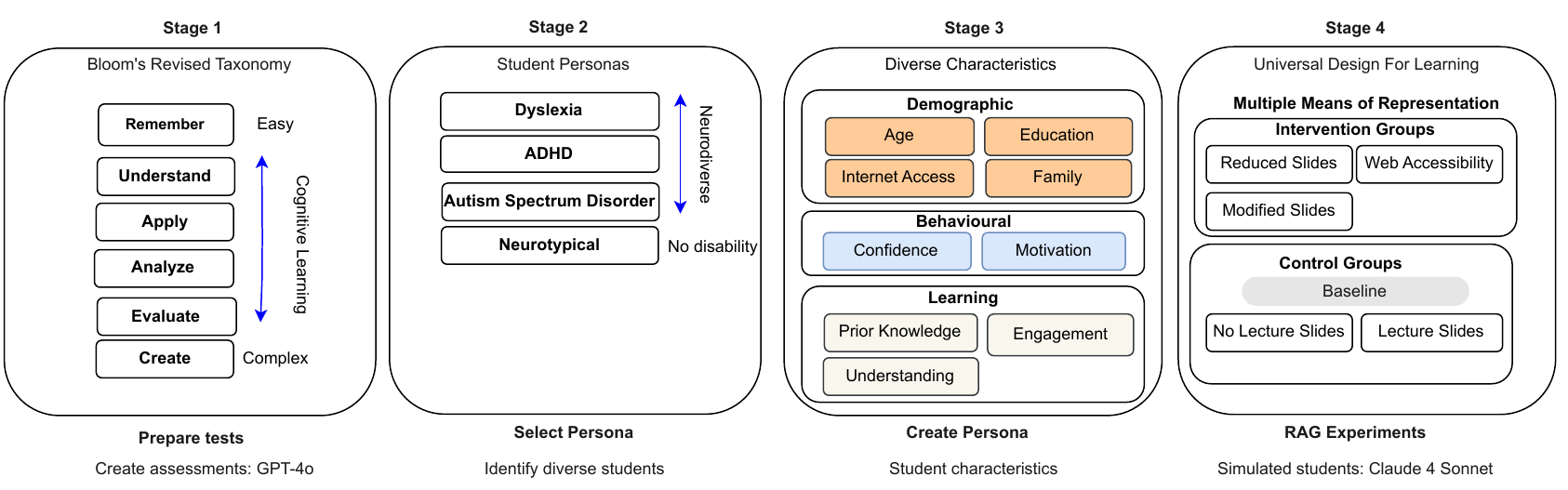}
    \label{fig:framework}
    \caption{DiverseClaire: An UDL Guided Approach to Improve CS1 Programming Lecture Slides For Accessibility}
    \Description{This is an image of the DiverseClaire UDL-Guided Approach}
\end{figure*}

The DiverseClaire pilot includes a UDL-guided approach to explore whether making introductory programming resources more accessible could benefit learners with different neurodiversity profiles. In Figure 1, the approach consists of four stages: a) design assessment questions using GPT-4o\footnote{https://openai.com/index/hello-gpt-4o/} b) select personas representative of a diverse university student population, c) create personas with diverse characteristics and d) experiments using Claude 4 Sonnet \footnote{https://www.anthropic.com/news/claude-4}, an LLM capable of generating text and code with reasoning capabilities.

 We designed randomised
 controlled experiments (RCTs)
 using students’ personas — representing
 dyslexia, ADHD, ASD and neurotypical
 learners to simulate how individuals might interact with course materials. Stanford University's 
lecture slides from the course 
\textit{CS106A: Programming Methodologies} in Spring 2025 (Week 3)
\footnote{https://tinyurl.com/Stanford-University-CS106A} served as our content base. We used GPT-4o an LLM supporting curriculum design \cite{Sridhar2023HarnessingObjectives} we produced assessment questions and solutions covering the six levels of Bloom’s revised taxonomy. 



To transform the slides into a more inclusive format, we applied UDL's Multiple Means of Representation and Web Content Accessibility Guidelines (WCAG) 2.2 in the DiverseClaire framework (Figure 1) to enhance lecture slides' inclusiveness. Following these principles, we
%
removed duplicate slides;
modified slide content by adding image alt-text, 
using sans-serif fonts for headings, and increasing font size to 28pt or higher; changed slides into an accessible Adobe PDF file format.

We generated synthetic students using few‑shot prompts with Claude 4 Sonnet based on our experiments comparing multiple versions of GPT, Gemini \footnote{https://tinyurl.com/LLM-Gemini-Pro-2} and Claude. Each persona combined demographic, behavioural and learning attributes  such as age, prior knowledge and engagement ~\cite{Xu2023LeveragingBehavior}. Three pedagogical factors were emphasised: prior knowledge, which provides context and improves the accuracy of LLM responses; knowledge background combined with engagement, which helps instructors tailor material; and behavioural or learning beyond demographics, since simulations based solely on demographic traits tend to fail.




Using an RCT design, we assigned 
experiments to a \textit{control group}
or \textit{intervention group} in Table 1. We applied RAG to generate responses and measured the intervention's effect. Each persona completed
five simulations. We evaluated LLM-based student responses against correct answers.The evaluation metric was the average score M. 
M represents the proportion of correctly answered questions across six levels of Bloom's revised taxonomy.

\vspace{-1 mm}

\section{Findings}

First, Stanford University's slides were presented in formats lacking digital
accessibility for neurodiverse learners \cite{Sahami2025CS106A25}.
To address this, we compared UDL-transformed slides against the original slides (Table 1).
The prediction accuracy of the simulated student suggests that with improvements to slides (Section 2), learners with 
dyslexia, ADHD, and neurotypical learners would be able to access slides
in a UDL-format with \textit{fewer slides and
web accessibility} and on average scored higher against control groups. 
Dyslexia and neurotypical learners
preferred \textit{fewer slides, modified content with web 
accessibility} and \textit{fewer slides, modified content without web
accessibility}. Both interventions
on average scored higher against the control groups. 
ADHD learners on average scored the highest
for \textit{lecture slides with web
accessibility} (M = 2.67), followed by identical scores (M = 2.50) for
\textit{fewer slides} and \textit{modified content}.
We discovered that UDL interventions to slides offered no benefit to ASD learners, with each intervention rated lower on average than the original slides (M = 2.50).
Neurotypical learners favoured \textit{fewer slides} (M = 5.00).

\begin{table}[]
\caption{RCT Results: Claude 4 Sonnet Student Simulations}
\scriptsize
\label{tab:my-table}
\begin{tabular}{lcccc}
\toprule
\multicolumn{1}{c}{Average Score (M)} & \multicolumn{4}{l}{} \\ \midrule
Scoring Method & \multicolumn{1}{l}{Dyslexia} & \multicolumn{1}{r}{ADHD} & \multicolumn{1}{l}{ASD} & \multicolumn{1}{l}{Neurotypical} \\
\textbf{Control Groups:} &  &  &  &  \\
Baseline I: no lecture slides & 0.00 & 0.00 & 0.33 & 1.33 \\
Baseline II: lecture slides + w/o accessibility & 0.67 & 2.00 & 2.50 & 4.17 \\
\textbf{Intervention Groups:} &  &  &  &  \\
modified content + w/o accessibility & 0.00 & 2.33 & 1.83 & 4.50 \\
lecture slides + accessibility & 0.50 & 2.67 & 1.83 & 4.83 \\
modified content + accessibility & 0.33 & 2.50 & 2.33 & 3.83 \\
reduced slides + w/o accessibility & 0.50 & 0.67 & 1.67 & 5.00 \\
reduced slides + modified content + w/o accessibility & 2.67 & 2.00 & 1.83 & 4.50 \\
reduced slides + accessibility & 1.83 & 2.50 & 2.17 & 4.83 \\
reduced slides + modified content + accessibility & 2.17 & 1.00 & 2.00 & 4.83 \\ \bottomrule
\end{tabular}
\vspace{-6mm}
\end{table}

\section{Conclusion}

The current research suggests that accessibility needs vary by 
learner. Neurotypical learners and those 
with dyslexia or ADHD 
preferred UDL-transformed slides, while learners with ASD did not.
These results highlight the potential
to offer CS1 course materials in multiple formats to accomodate different learner 
preferences. 


\bibliographystyle{ACM-Reference-Format}
\bibliography{mysigreferences}

\end{document}